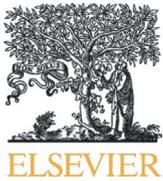

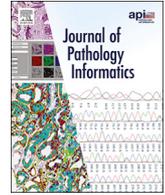

# Iris: A Next Generation Digital Pathology Rendering Engine

Ryan Erik Landvater *, Ulysses Balis

*University of Michigan Medical School, Department of Pathology, 2800 Plymouth Road, Ann Arbor, MI 48109-2800, USA*

| ARTICLE INFO | ABSTRACT |
|---|---|

Digital pathology is a tool of rapidly evolving importance within the discipline of pathology. Whole slide imaging promises numerous advantages; however, adoption is limited by challenges in ease of use and speed of high-quality image rendering relative to the simplicity and visual quality of glass slides. Herein, we introduce Iris, a new high-performance digital pathology rendering system. Specifically, we outline and detail the performance metrics of Iris Core, the core rendering engine technology. Iris Core comprises machine code modules written from the ground up in C++ and using Vulkan, a low-level and low-overhead cross-platform graphical processing unit application program interface, and our novel rapid tile buffering algorithms. We provide a detailed explanation of Iris Core's system architecture, including the stateless isolation of core processes, interprocess communication paradigms, and explicit synchronization paradigms that provide powerful control over the graphical processing unit. Iris Core achieves slide rendering at the sustained maximum frame rate on all tested platforms (120 FPS) and buffers an entire new slide field of view, without overlapping pixels, in 10 ms with enhanced detail in 30 ms. Further, it is able to buffer and compute high-fidelity reduction-enhancements for viewing low-power cytology with increased visual quality at a rate of 100–160 µs per slide tile, and with a cumulative median buffering rate of 1.36 GB of decompressed image data per second. This buffering rate allows for an entirely new field of view to be fully buffered and rendered in less than a single monitor refresh on a standard display, and high detail features within 2–3 monitor refresh frames. These metrics far exceed previously published specifications, beyond an order of magnitude in some contexts. The system shows no slowing with high use loads, but rather increases performance due to graphical processing unit cache control mechanisms and is "future-proof" due to near unlimited parallel scalability.

## Introduction

As a preface, it should be stated from the outset that the viewing of whole slide images (WSIs) firmly qualifies as a high-performance computing venture.[1,2] Despite this reality, both vendor deployments (SectraAB and Aperio Webviewer[3]) and academically based explorations[3–10] alike have generally approached the development of workflow solutions from the confines of traditional thin-client development stacks, carrying with those efforts the inherent and usually insurmountable limitations, as imposed by single- or oligo-threaded architectures without integrated graphical processing unit (GPU) compute capabilities.

Conversely, the progression of central processing unit (CPU) hardware acceleration, GPU parallelization technologies and massive local memory have the combined potential to revolutionize the digital pathology viewing experience for primary diagnosis. Many current whole slide viewer (WSV) software packages, such as those based on the OpenSeadragon library,[4,8,11] make use of the JavaScript frameworks,[7] which operates as an essentially interpretive (non-compiled) application. These strategies interact with the GPU using the WebGL application programming interface (API), a stateful non-parallel graphics library. Alternative academic implementations use adapted generic image rendering technologies based upon the .NET[5,10] or Qt frameworks,[6,9] which possess similar limitations as WebGL. The reduced speed and data throughput of such frameworks are neither optimized nor ideal for this scale and complexity, as represented by the volume of data intrinsic to digital pathology workflow in primary diagnosis settings. Limitations of malalignment between the core rendering technology and the end-use requirements are as follows:

Performance bottlenecks: Traditional JavaScript frameworks can suffer from performance bottlenecks when handling extremely large images, which are common in digital pathology. Interpretive applications are converted to machine code from human-readable code at the time of use with reductions in optimization such as single instruction multiple data (SIMD; parallel computation) instructions. This often manifests as rendering lag in both image tile retrieval speed and tile resolution refinement.

Inefficiency in data handling: Inefficient data handling and rendering processes can lead to delays in image loading and manipulation, which can similarly impact the pathologist's workflow.

* Corresponding author.
*E-mail address:* ryanlandvater@gmail.com (R.E. Landvater).

# Iris System Architecture and API Overview

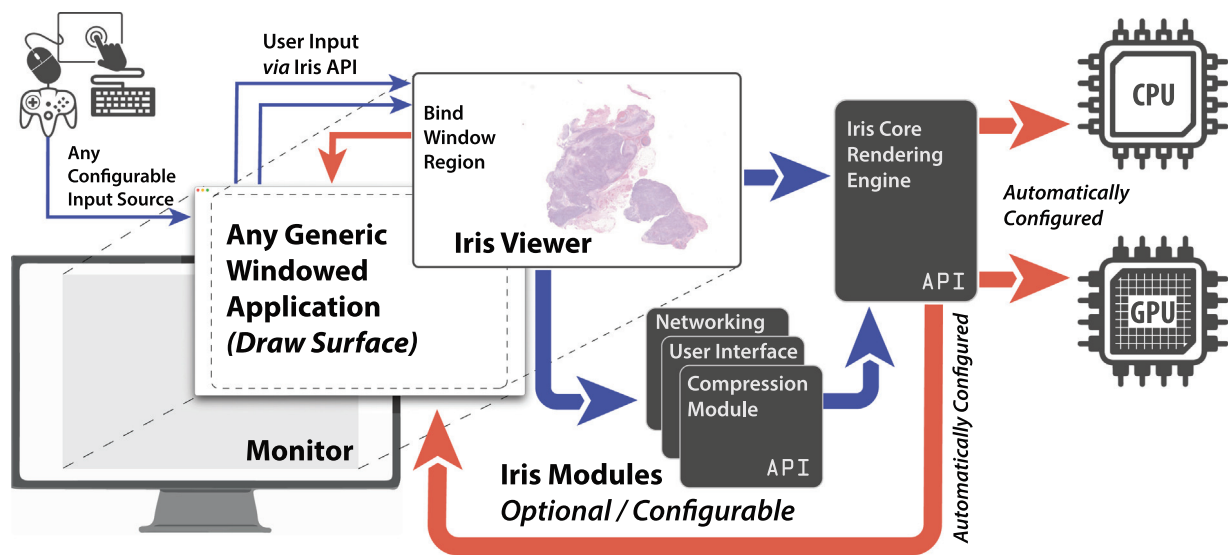

**Fig. 1.** Iris implementation and application programming interface overview. The Iris system is composed of multiple modules, which can optionally be configured to interact with Iris Core, the WSI-specific and optimized rendering engine. Iris Core is compiled for Windows, macOS, iOS, and Linux from portable C + + source code and can bind the draw-surface of a generic WSV application's graphical window to begin drawing slide data. Iris's API (blue arrows) is accessed through the Iris Viewer instance, which coordinates the active modules and renders slide data to the bound window. The system automatically configures runtime parameters (red arrows) based upon identified hardware capabilities and the calling application's operating system window visual surface, a feature known as a "plug and play" configuration. The Iris Core API is lightweight comprising only a few header files.

Limited parallel processing: Most of these frameworks do not fully exploit modern CPU and GPU capabilities, resulting in suboptimal performance, below what is theoretically possible for a given hardware platform. This is especially evident when viewing high-resolution (HR), multi-gigapixel images.

Super-pipeline architectures[12,13] are an emerging solution designed to maximize the utilization of these advanced CPUs and GPUs. These architectures are increasingly utilized in graphical programming at a critical time when the field of pathology shifts from glass slides to digital sign-out. Systems designed with super-pipeline architectures benefit from massive parallelization and demonstrate improved screen refresh latencies capable of operating at the current hardware limits. They are designed with future-proof scalability in mind, such that an indefinite degree of hardware parallelization can be immediately utilized to its fullest potential. In this technical communication, we report on one example reference architecture (Iris), which is an optimized and modular WSI rendering system that can be directly integrated into a variety of WSV platforms and that fully leverages the potential of massive hardware parallelization and super-pipelined architectures.

## Design and implementation

Iris is a series of portable machine-code WSI library modules developed over a 3-year period for use in creating or enhancing WSV applications or solutions. This communication focuses on the rendering engine module of Iris, also referred to as "Iris Core." Iris Core is implemented with a light header file, in addition to the machine code library, in a "plug-and-play" manner (the system will automatically configure hardware runtime parameters with limited input required by the calling processes; Fig. 1, *red arrows*). The system is written primarily in C + + and compiled for Unix-based (macOS, iOS, Linux) and Windows operating systems. Iris' rendering engine is built to support Vulkan, the low-level, low-overhead cross-platform API and open standard for 3D graphics/GPU computing created and maintained by the Kronos Group (Beaverton, OR). In addition to Vulkan, Iris Core has limited other dependencies, such as graphical decoders (libPNG and TurboJPEG), and text rasterizers (FreeType). Iris may render WSI files in any format supported by OpenSlide (OS),[14] an open-source vendor-neutral WSI decoder[1] and may optionally encode any such file into the Iris Codec format (.iris) for faster rendering. The system can be implemented in only a few lines of code.[2] The random-access memory (RAM) and GPU's video RAM (VRAM) usage can be configured at runtime but generally requires little more than 2 GB combined. Iris is written with a compact codebase to create a secure system with minimal fail-points and attack surfaces while keeping efficiency as a paramount characteristic. The system architecture implementation (Fig. 1) is described below.

Iris draws the WSI using tile-rendering techniques, wherein WSI dimensions are divided into several 256 × 256-pixel tiles based on the image layer resolution, and only the immediately relevant tiles are rendered. The choice of 256-pixel tiles was made for ease of GPU bit operations and for ongoing work related to digital compression, which uses 16-pixel sub-tiles. The scope view, in which the WSI tiles are rendered, is a multi-pass rendering sequence that can include stenciling buffers for increased efficiency. The number of tiles within the scope view is a function of the screen size and resolution. The two drawing passes correspond to two resolution layers: LR and HR and are detailed in Fig. 2. The low-resolution (LR) sub-pass is stenciled and corresponds with the WSI layer whose 256-pixel tiles, based upon the current zoom level, would be drawn to occupy more than 256 pixels of the screen (1:1 image-to-screen pixel ratio); that is to say, this layer represents the tiles that are enlarged relative to their native resolution. The HR layer is drawn "above" the LR layer and represents tiles with a greater resolution than the screen resolution for the current zoom level (≤ 1:1 image-to-screen pixel ratio). These tiles are shrunken and drawn smaller than their native resolution (using mipmaps[15]); thus, they contain more image information and detail. Any HR tiles that are not yet buffered are treated as transparent. Within these mosaic regions of the scope view, the underlying LR tiles are thus visible through the transparent HR tile regions until the HR tiles corresponding to those regions complete their buffering.

---

[1] Please refer to the OS documentation for supported WSI file formats.
[2] Please refer to the IrisExamples Github repository for example implementations.

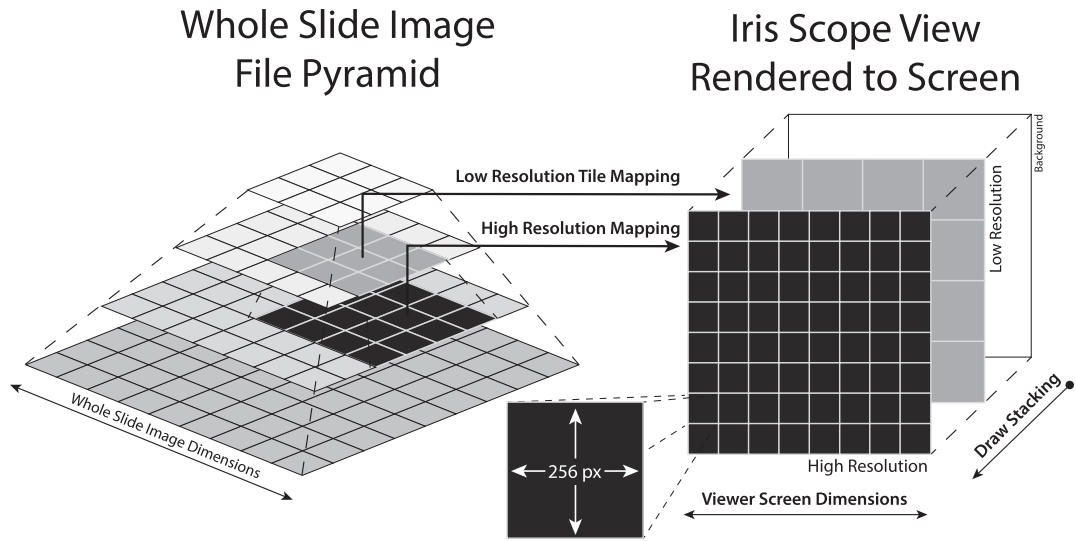

**Fig. 2.** Whole slide image (WSI) to visible draw-space tile mapping scheme. The WSI layer pyramid (left), comprising multiple sub-sampled image layers at decreasing resolution (zoom-level), are mapped to visible tiles drawn to the screen as part of Iris' rendering pipelines (right). These visible tiles represent 256 × 256-pixel regions of the WSI pyramid layers, and the number of tiles rendered are equivalent to the screen dimension, divided by 256 pixels, and multiplied by the current zoom factor for that layer. The high-resolution layer represents ≤1.0 WSI pixel to monitor pixel (i.e., these tiles are shrunken to less than their native size), whereas the LR layer represents 1.0 WSI pixel per monitor pixel (i.e., these tiles are enlarged to greater than their native size). Any unbuffered tile space is transparent, such that aspects of the LR layer are visible during the high-resolution buffering.[16]

## CPU architecture

Iris Core engine's architecture involves numerous asynchronous processes operating on the CPU and GPU. We will begin by detailing the CPU elements. Iris Core operates numerous concurrent threads on the CPU (Fig. 3, *top*). One or more user input threads external to Iris may provide calls that update Iris' state, such as opening or closing a slide, slide translations, adjusting the view zoom/objective, etc. Calls to Iris' Core API are thread-safe. The main CPU concurrent elements are as follows: (1) a rendering thread, which issues scope-view draw commands; (2) a buffering thread, which operates hierarchical transfer queues and issues buffering commands; (3) numerous image loader threads, which read slide tiles from the WSI file, and (4) a series of asynchronous callback threads, which perform numerous system functions (not illustrated in Fig. 3). Coordination between the various threads occurs via atomics (i.e., thread-safe and non-blocking variables) within resource directories. The two main resource directories are: (1) a scope resolution-layer GPU-resource map (Fig. 4) that informs the system which GPU texture resources correspond to a particular combination of slide location and resolution layer and (2) a short-term image cache, where decompressed image tiles are cached in RAM for use by the buffering thread queues both in regions of rendering and an additional outside buffering perimeter (usually a 1–3 tile radius).

Iris treats each CPU element as an independent process with interprocess communication protocols without thread blocking. Iris' asynchronous interactions are governed by a pull-style/hybrid architecture, in that a slide image is not waited upon but rather pulled from a common resource directory when signaled via interprocesses communication. This is in contrast to a push architecture wherein an upstream process "pushes" data, which may become stale, to a downstream process (for example, a loader thread initiating the buffer thread transaction with a list of tile it has made available). Iris' processes constantly fetch and generate their own models of the current view to avoid stale data and avoid waiting on upstream processes. While this type of architectural paradigm is typically used in networking where server and client machines are independent and asynchronous, independent CPU process handing make it well suited to Iris' architecture. Furthermore, it integrates well with rendering APIs as the separate CPU–host and delegate GPU–device paradigm is very similar to networking request–response interactions. This architecture was adapted for use in Iris due to the highly dynamic nature of high-speed scope rendering and stateless design of each CPU process. The region that the scope view renders shifts often; as a result tile indices may become stale, even between requesting and rendering—for example, during very rapid saccade-like slide movement. The solution Iris utilizes are interprocess request protocols for resources made via lock-less communication paradigms which are never waited upon. This forces each CPU thread to operate in a stateless manner with its own model of the slide view regions. Threads operate without pausing and cycle until there are no more tasks to operate upon. When a process completes a task, it send a generic interprocess signal to other processes that a change may have occurred and to recalculate models to re-assess for possible work. These relationships are shown in Fig. 3, top frame.

## GPU pipeline architecture

Each process described above issues commands to respective GPU queues concurrently and we will now describe how these commands are carried out by the GPU pipelines, illustrated in Fig. 3, *bottom pane*. Iris configures internal parameters based upon queried GPU capabilities during execution but optimally operates three separate queue families. Queue families are GPU data-paths with unique capabilities that can operate simultaneously with respect to other queue families' operations. Concurrent queue family submissions have essentially no effect on one another's parallel performance; however, sharing access to a single queue for multiple tasks has marked performance implications, similar to the performance of older APIs like WebGL/OpenGL. This unique stateless feature of modern APIs like Vulkan allow Iris to issue GPU commands from multiple CPU threads simultaneously, agnostic of each other's submission statuses. The buffer thread schedules three primary commands (Fig. 3, *center-right bottom pane*): (1) bring slide tile data from staging mapped GPU memory into the high-performance GPU-local memory; (2) enhance using the single pass downsample compute pipeline (a unique feature of Iris that enhances cytological features when zoomed out; detailed in Fig. 5); and (3) mark the image as visible within the respective resolution-layer. The specifics of this process, the rapid tile buffering sequence (RTBS), will be described in greater detail later. (See Fig. 6.)

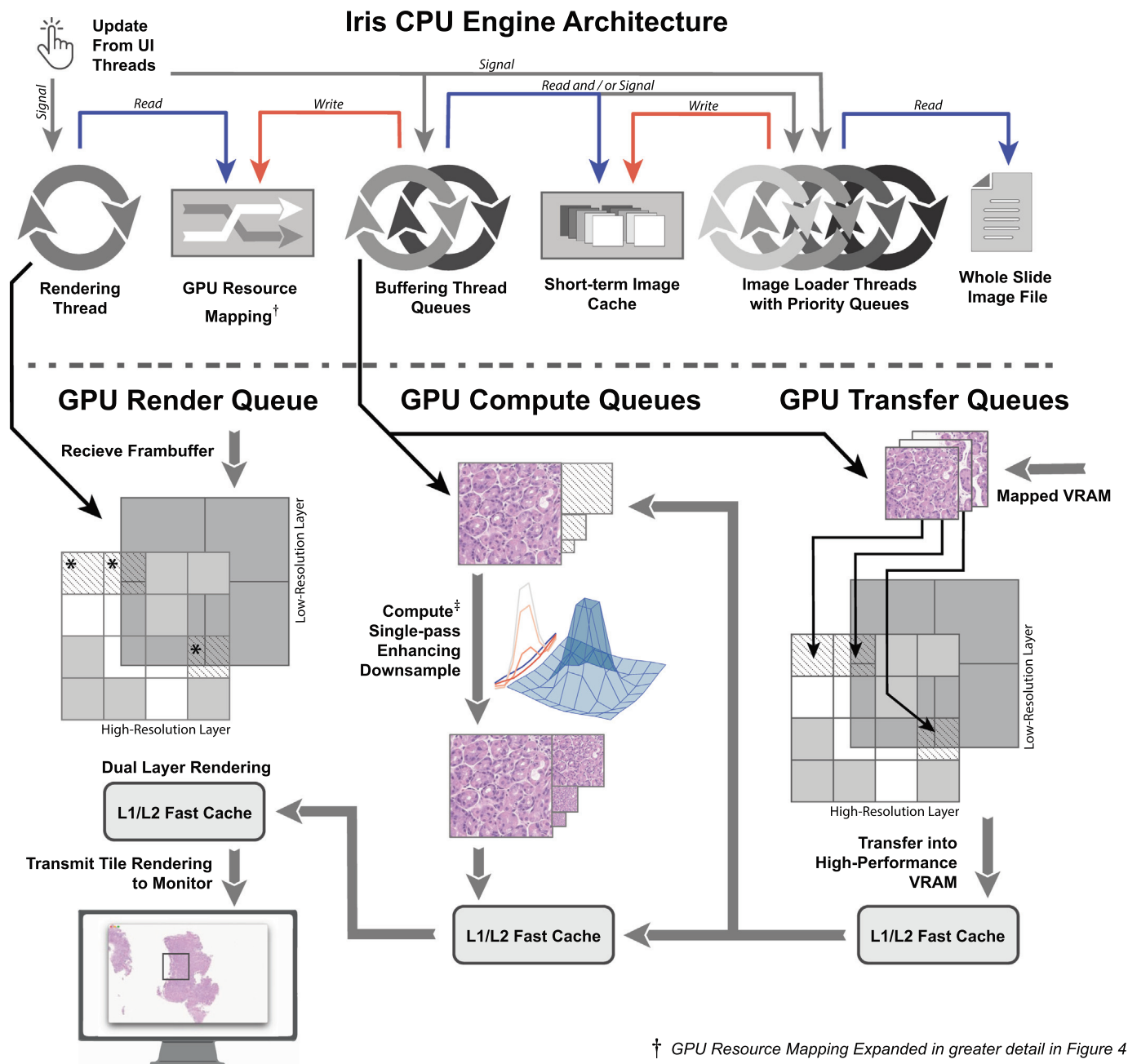

**Fig. 3.** Iris system architecture detailing CPU and GPU-pipelines. CPU engine architecture (top) involves coordination between multiple concurrent thread executions with simultaneous GPU submissions (arrows that cross to bottom). The user interface (UI) threads signal the rendering, buffering, and loading threads of region updates. The rendering thread pulls GPU resources (blue arrow) to render via an atomically updated VRAM-to-slide-location map[†] (Fig. 4), updated in real-time by the buffer thread (red arrow). The buffer thread executes lock-less high and low-priority queues and stacks by pulling (blue arrow) raw image data from a short-term RAM cache and initiating GPU transfer and compute pipelines. Additionally, the buffering thread queues and prioritizes concurrent loader threads that read slide data from a slide file or locally cached server slide-file (blue arrow). Multiple concurrent GPU queue families (bottom) execute render and buffer commands, all of which may be submitted simultaneously. Transfer commands and enhance/downsample pipeline[‡] (Fig. 5) are executed in series with synchronization of the L1/L2 GPU cache between queues. Actively buffering tiles (*) may be rendered during a render-pass if the transfer completes before fragment shader execution and control of that location within the L1/L2 cache is transferred over to the render queue.

Simultaneously and independently, the rendering thread checks if the view region has changed and issues draw commands for any fully buffered visible tiles of both resolution layers within the scope view region. Because execution occurs concurrently within multiple queues/queue families and because the rendering of a tile frequently occurs before buffered data have returned from the GPU cores to VRAM, direct control of tile image data within the GPU cache must be explicitly transferred between buffer, compute, and render queue families. The interplay between buffering and rendering occurs so quickly that failure to explicitly transfer GPU cache control will cause visual artifacts (further detailed in Fig. 4, *caption*). A newly buffered tile, denoted by an asterisk (*) in Fig. 3, may be rendered during a render-pass if the buffer command completes before the render pass fragment shader (the GPU render stage during which the image tile pixels are drawn to the scope's view), and cache control of those tiles was given to the render queue.

## Coordination of Simultaneous Read-Write Commands

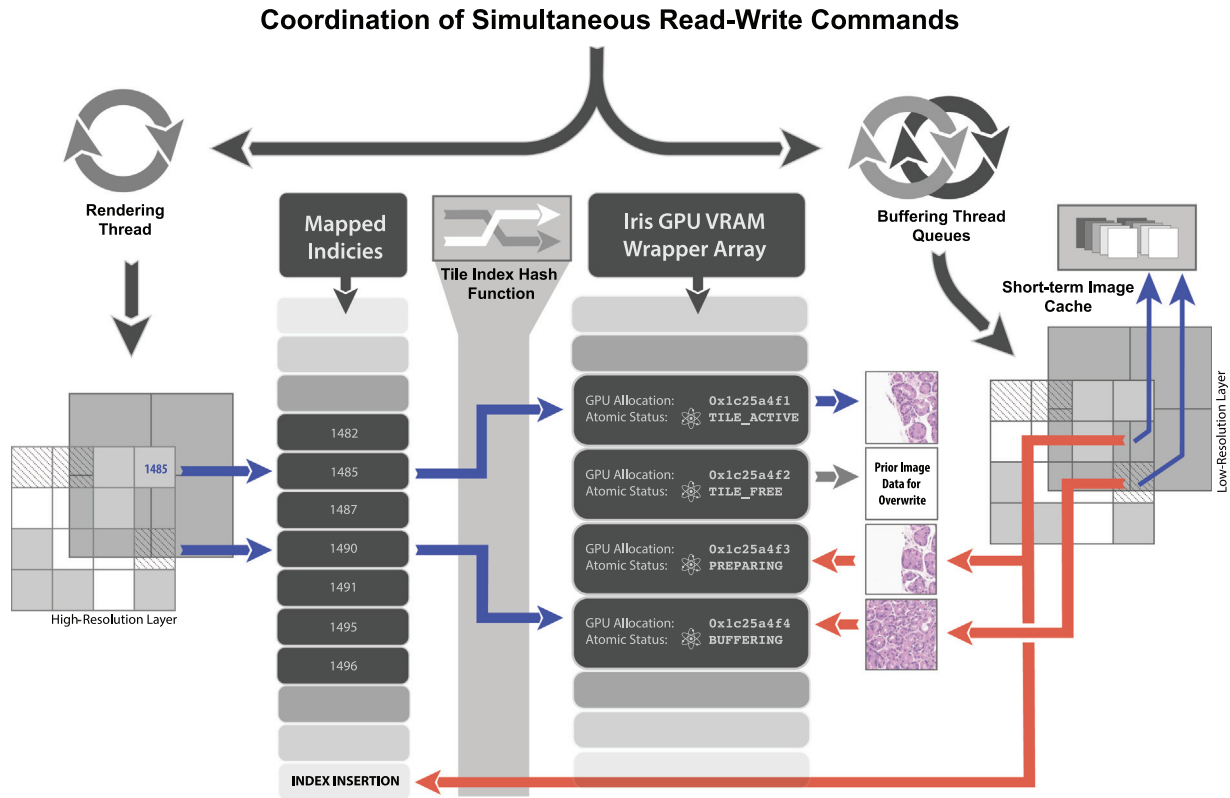

**Fig. 4.** Simultaneous buffer-render coordination schema. Physical tile locations with index values (left) are hash-mapped to respective tile image data residing within associated GPU VRAM allocations. VRAM image allocations are also indexed within an Iris VRAM Allocation Wrapper array. Allocations are recycled and their image data is overwritten to avoid GPU allocation overhead. Atomic reference counting and atomic status flags are associated with each Iris Wrapper Instance to track use across concurrent threads. Read commands, illustrated in blue, show how the tile index is mapped to the proper index within Iris Wrapper Instance Array for drawing the image allocation associated with active tiles (1485 for example). An indeterminate status for outstanding tiles may exist as well based upon GPU buffering progression (1490). Image data are read from a dynamic short-term cache and write commands, denoted in red, transfer image data into available (status TILE_FREE) allocations during microtransactions. These allocations are identified when the thread queries the Iris Wrapper Instance array for available allocation wrapper instances or purges the state of mapped tiles away from the rendering view. Allocation purging simply requires flipping the atomic flag from TILE_ACTIVE to TILE_FREE as buffering overwrites the prior image data. This simple schema makes image deallocation or clearing unnecessary, however it does require proper GPU L2 cache control mechanisms to avoid rendering stale data.

### Rapid tile buffering sequence

The core technologies that allow for Iris to render WSI data quickly are referred to as the RTBS and are described in greater detail within the patent "Technologies for Improved Whole Slide Imaging".[16] The RTBS follows super-pipelining paradigms and divides the process of buffering view tiles into numerous hierarchically prioritized microbuffering transactions, during which only a small number of immediately available tiles may be buffered in a single call. The "pull"-style architecture ensures any newly available tiles will be enqueued for buffering as soon as possible, without waiting on additional data. The plurality of microtransactions allows the GPU scheduler to take full advantage of super-pipelining, staggering transaction to saturate GPU cores, which reduce the amount of time to buffer each tile as a greater numbers of tiles are added. This is illustrated in the results section with HR layers buffered faster per tile (Fig. 7). LR layer data are prioritized to ensure at least a LR view is available, whereas the HR details are buffered in a piece-wise manner during subsequent pipeline passes. A buffer region (the buffering model), defined as the view-space plus some arbitrary perimeter, is statelessly re-calculated by the buffering thread during each pass to ensure only the most relevant tiles are sent to the GPU dynamically. The buffering thread queries the GPU resource map (Fig. 4) to assess which relevant tiles are already mapped, transferring, or available for buffering. Any unavailable tiles are then prioritized by the loader thread queues and any available unmapped tiles are immediately sent to the GPU queue for buffering. Each microbuffering transaction occurs rapidly, such that more than one may be submitted and enqueued for GPU execution per frame draw call. Thus, the GPU will execute multiples of these microbuffer transactions simultaneously but staggered by the GPU scheduler (super-pipelining) with each draw command. A subset of these will complete during the draw process and be rendered. Explicit synchronization of GPU and CPU resources is a challenging but critical aspect of the RTBS due to the immense data volume and speed of data transactions within the vast number of microtransactions that span draw calls as well as shifting zoom levels and view planes. Queue synchronization involves a highly coordinated set of atomic variables and on-GPU synchronization tools that are beyond the scope of this communication.

### Single pass reduction-enhancement

Another salient feature of the engine design pertains to scope image quality. Unlike a physical microscope with objectives, digital microscopes can render continuous zoom levels. Mipmaps are generated for this reason. Mipmaps are pre-calculated progressively smaller versions of the original image ($2\times$, $4\times$, $8\times$, etc.). Tiles are drawn via texture sampling, in which GPU image data within VRAM allocations are read and mathematically altered to best fit the physical display space. This may involve linear algebra functions such as weighted blending operations between mipmap levels when the tile dimensions (256 pixels) do not exactly match the dimensions of the rendered tile, due to zooming in or out, which is the case the vast majority of the time. Without use of mipmaps, sampling quality is severely reduced with visual artifacts such as Moiré patterns.[15] Iris Core includes a specialized GPU compute pipeline to compute mipmaps (Fig. 3,

## Single Pass Reduction-Enchancement Downpass Compute Pipeline

### Wide GPU SIMD instruction sets compute all MIP levels on all tiles, concurrently

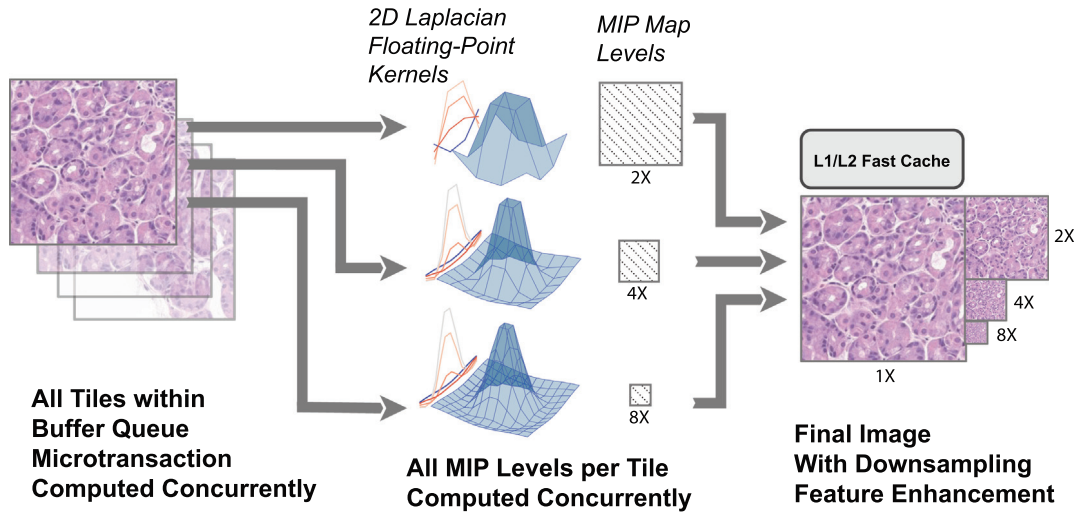

**Fig. 5.** Single pass mipmap generation compute pipeline schema. During a buffering microtransaction, all tiles (left) within the buffering microtransaction are enqueued for mipmap generation. Each mipmap for a given tile is generated simultaneously from the reference tile ($1\times$) that was buffered to the GPU using a single pipeline pass, a process known as single pass downsampling.[17] Variably sized and normalized 2D Laplacian sharpening kernels (using larger kernels for greater reduction), comprising floating-point numerals, are convolved on the reference image for each mipmap level ($2\times$, $4\times$, $8\times$, etc.) to generate images of the highest possible visual quality. The resultant tiles demonstrate enhanced cytological detail when sampled for rendering at lower magnification power.

*GPU middle column*; Fig. 5) with three salient features: (1) All mipmap levels are generated in a single GPU pass for all tiles of the buffering microtransaction at once, known as a single pass downsampler (SPD)[17]; (2) Mipmap generating calculations are performed with floating point kernel convolutions to avoid integer quantization error and are generated only from the $1\times$ reference image to avoid cumulative error (which is common with mipmap generation due to successive reduction of levels); (3) Multiple 2D Laplacian (of Gaussian) kernels are used for reduction-enhancement to avoid the blurring of important cytological details. When mixing pixels in the generation of smaller mipmap images, averaging of these pixel values over the edges of nuclei or cytoplasmic elements can cause blurring of these salient features. This last point is critical to low-powered visualization of high-value features such as mitotic activity or chromatin character. The method used in Iris is similar to the feature enhancement of the human eye via lateral inhibition by the horizontal and amacrine cells.[18] This pipeline, outlined visually in Fig. 5, is designed specifically for digital slides in the Iris Core engine but uses a similar implementation to that of AMD's advanced rendering research group's FidelityFX SPD.[17]

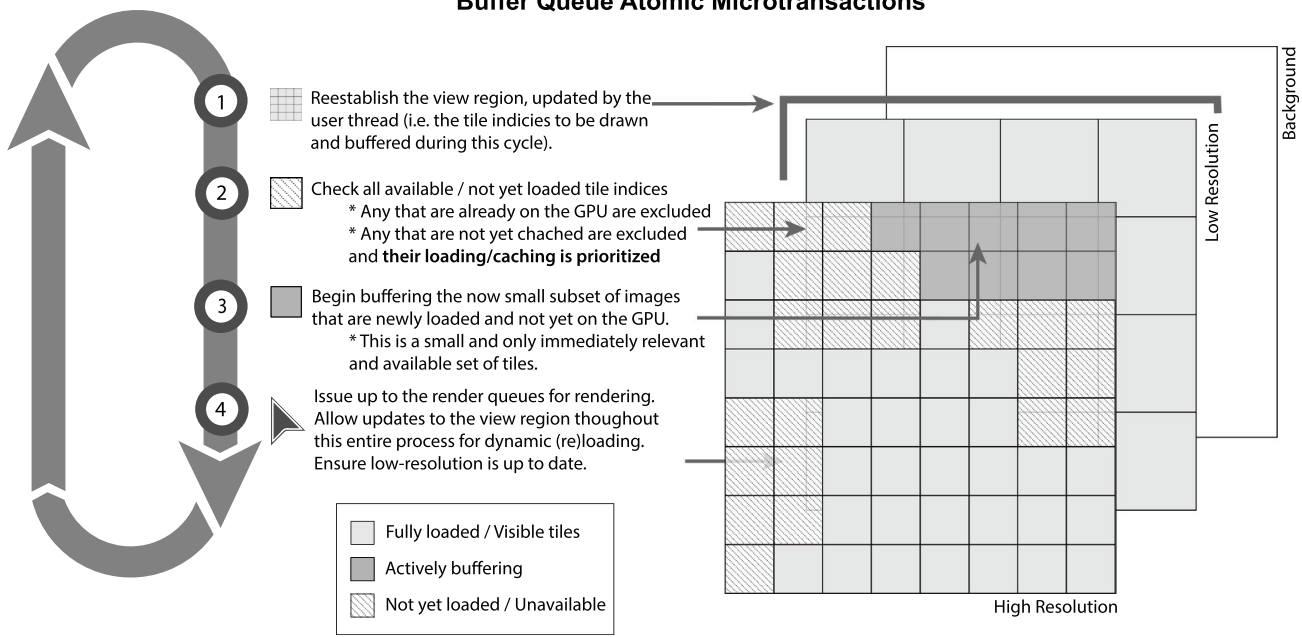

**Fig. 6.** Rapid tile buffering sequence (RTBS) atomic microtransactions. The RTBS breaks pending buffering transactions into numerous microtransactions (which utilize atomic state signals for concurrency and GPU device synchronization) and coordinates GPU resource recycling to avoid allocation overhead. Using the described "pull" design principles, the transactions only buffer data immediately available in regions pending buffering. This results in multiple microtransactions per frame and the GPU scheduler staggering transaction to saturate GPU cores. Unavailable data are prioritized by the loader threads and unbuffered tiles are made transparent so that only the LR regions are rendered while the buffering completes the high-detail sections.[16]

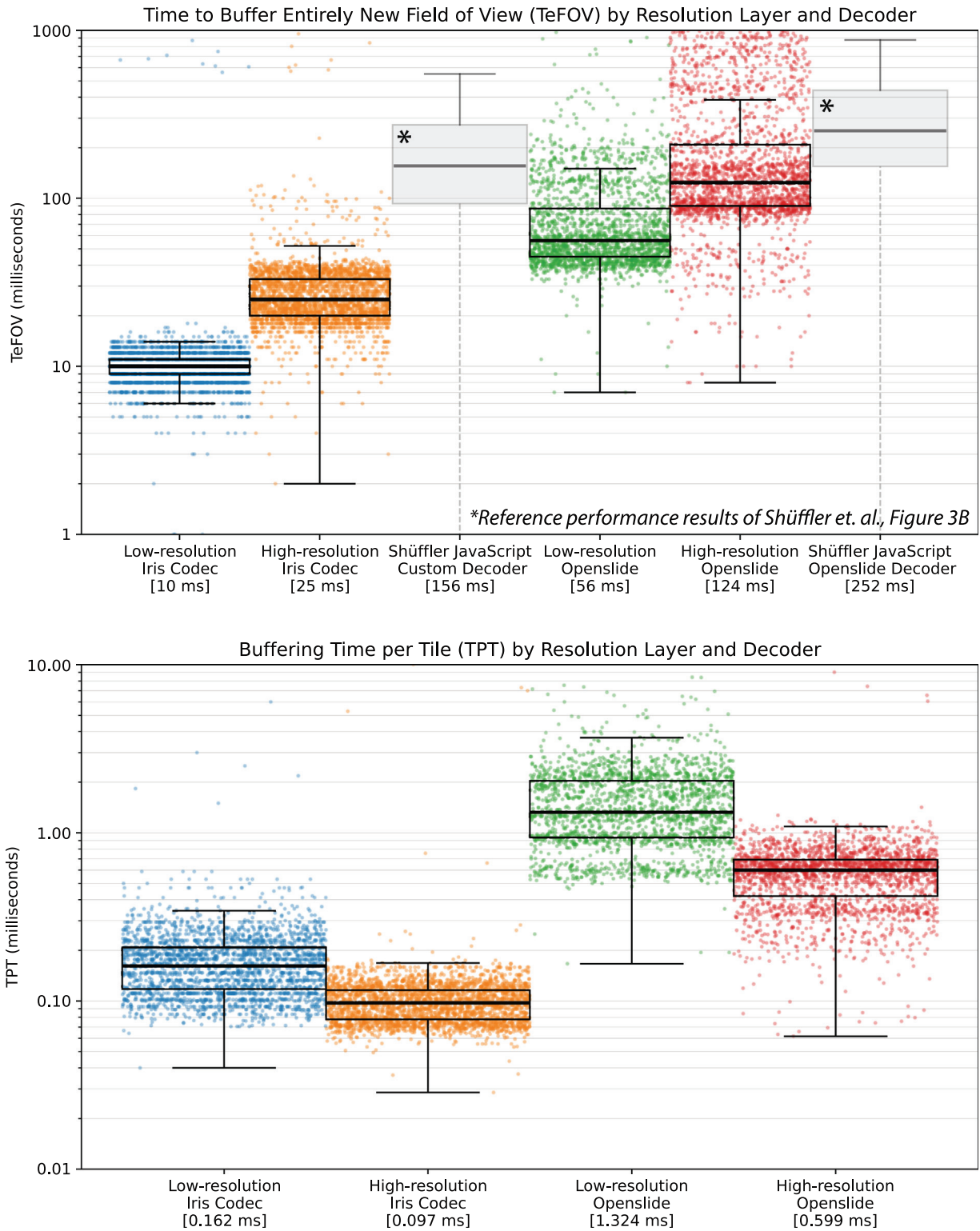

**Fig. 7.** Iris buffering rate for time for field of view (TeFOV) and time per tile (TPT). Scatter plots of individual buffering events and boxplots showing aggregate TeFOV (top) and TPT (bottom), plotted base-10 logarithmically to show order of magnitude performance differences. Median TFOV times decoding using Iris Codec of LR-FOV 10 ms and HR-FOV 25 ms; when decoding with OpenSlide LR-FOV 56 ms and HR-FOV 124 ms. These TFOV times are shown in the context of previously published high-performance JavaScript by Schuffler *et al.* denoted by asterisk (*).[7] TPT values decoding using Iris Codec were LR-TPT 160 µs and HR-TPT 100 µs; when decoding with OpenSlide LR-TPT 1320 µs and HR-TPT 600 µs.

## Performance results

Iris Core is written with performance and ease-of-use as the paramount qualities during development. Performance is broken into two facets that must independently be considered, but which impact one another based on proper system design: (1) Frame-rate/user responsiveness and (2) WSI data buffer-rate. Frame-rate and user responsiveness define how rapidly the system updates the image on the screen and how quickly the images change in response to user input. WSI data buffer-rate determines how quickly the underlying image data is transferred to the GPU to be rendered to the screen. With low frame-rate, the system responds slowly with jerky movements. Slow WSI buffering causes the system to move smoothly but renders blurred or incomplete slide images.

All following metrics are based upon example implementations of Iris Core (which is natively a tool to build WSV applications). These example implementations were written for macOS/iOS using Xcode in Objective C + + and are made available as source code (see supplement). To evaluate the WSI data buffering speeds, the data-transfer rates of the buffer thread queues were evaluated in terms of actual raw data in gigabytes per second (buffer-rate) as well as tiles per millisecond (tile-transfer rate) and functionally in the form of milliseconds needed to buffer a scope field of view (buffer-time). The frame-rate was also used to establish a consistent time interval for calculating the raw data buffer-rate and to allow for a constant time based x-axis when co-plotting these buffer- and frame-rate tracings. The following empiric parameters were defined for this cause and partially based on the work by Shüffler *et al.*[7,8]

**Frame-rate:** The number of scope-view image frames drawn per second, calculated as the inverse of time between drawn frames. This was also used as the frequency with which the data were collected for use-tracing and the buffer-rate, as it provided a relatively consistent and relevant interval by which to collect system performance data.

**Buffer-rate:** The amount of data the GPU transferred into high-performance memory and enhanced by single pass downsampling per approximate fixed-time interval, based upon frame rate. The buffer-rate value is defined as the sum of the bytes the GPU reported it had completed transferring (and enhancing) between the start of a frame and the prior frame divided by the duration of that frame in seconds (the inverse of the frame-rate).

**Time to entirely new field of view** (TeFOV): The amount of time to buffer the tiles that comprise an entire screen field of view (FOV), specifically the time between the user signaling a "move-field" request via Iris' API and the buffering of all tiles within an entirely new field. This differs from the TFOV, defined in the work by Shüffler *et al.*, in that a new field is explicitly defined as a scope-view without any overlapping pixels with the previous field of view. If there was any situation in which a new field was already buffered, due to accidental re-tracing of a previous region, those data points were excluded.

**Time per tile** (TPT): The average amount of time to buffer each $256 \times 256$-pixel tile per new field of view buffering process. This is equivalent to the TeFOV divided by the number of titles buffered within the FOV and the additional outside perimeter (usually a 1–3 tile radius).

### TeFOV and TPT

TeFOV and TPT performance characteristics were evaluated on a 2020 13-in. M1 MacBook Pro (Apple, Palo Alto, CA) with 8 GB of RAM (average Iris RAM/VRAM consumption was 2.5 GB sustained) with a window resolution of $2774 \times 1750$ pixels (4 k retina) running macOS 13.5 and using MoltenVK as the GPU driver library for Vulkan implementation. WSI files were Lecia (Leica Biosystems, Buffalo Grove, Illinois, USA) SVS files, read with OS,[14] an open-source vendor-neutral WSI decoder. The same files were converted to Iris Codec file format (.*iris*) and read using the Iris Codec (Iris Compression Module). The median times, with 25–75th percentiles in brackets, to render an entirely new LR FOV with Iris Codec was 10 ms (9–11 ms) and HR FOV with Iris Codec was 25 ms (20–33 ms). The time to render an entirely new LR-FOV when decoding using OS was 56 ms (45–87 ms) and a new HR-FOV using OS was 124 ms (90–209 ms). The TPT rendering when using Iris Codec for decoding LR tiles (LR-TPT) was 0.16 ms per tile (0.12–0.21 ms) and for HR tiles decoding was 0.10 ms per tile (0.08–0.12 ms). When rendering using OS for decoding, the LR-TPT was 1.32 ms per tile (0.94–2.04 ms) and HR-TPT using OS was 0.60 ms per tile (0.42–0.69 ms). These results, co-plotted with reference values from Shuffler *et al.* (2022), are displayed in Fig. 7.

### Frame-rate and buffer-rate

Iris has achieved the sustained maximum frame-rate (which we defined as the median frame-rate for rendering equal to the maximum monitor refresh rate) on all tested platforms. We conducted frame- and buffer-rate tests on an Apple M1 11in. 3rd generation iPad Pro (Apple, Palo Alto, CA) with resolution $2420 \times 1668$ pixels (4 k retina; 120 Hz refresh) running iOS 17 and recorded data using Xcode's Metal performance monitor as well as Vulkan's Query API. MoltenVK was again used as the driver library. This platform was chosen for the combined FPS-buffer tracings due to the 120 Hz monitor, which is twice the standard desktop monitor refresh, responsive control of touch-screen interface, and for the additional benefit of integrated Apple Pencil annotations which requires co-rendering with a separate external rendering pipeline (Apple, Palo Alto, CA). Results from a 12-s use-tracing, sampled every 7–9 ms, of an Iris Codec slide file are shown as a representative example (Fig. 8). All other test tracing results (not shown) were nearly identical. Use traces showed a sustained 120 FPS (119.4–121.9 FPS) frame-rate with rapid slide movements (saccade-like movements) like that of manual glass-slide driving using the touchscreen as the user interface driver for slide navigation (see supplemental Video 1). Image data were transferred and downsampled in short bursts that correspond with user slide movement. These buffering bursts of slide data occurred at rates of 1–2 GB of decompressed image data per second (median 1.39 GB/s (1.01–1.69 GB/s)). Rendering shader execution times were ≤2 ms and not prolonged by high concurrent buffering GPU load; rather, transient but significant reductions in shader execution times were noted to correlate with large data bursts (Fig. 8; *red arrows*) due to GPU caching operations.

There is often a minor decrease in frame-rate followed by spike and then similar decrease (trough–peak–trough patterns) seen on the frame-rate trace (Fig. 8, *top*). The first trough coincides with the beginning of buffering transactions and is due to the short burst of maximal GPU workload. This frame-rate reduction quickly disappeared as the shortened shader execution times overcompensate for the increased GPU hardware demand. The peak is the result of decreased shader execution time and brings the rendering pipelines of the engine ahead of the monitor's refresh rate. This requires a compensatory decrease in frame-rate to resynchronize with the monitor, representing the second FPS trough. These trace results demonstrably illustrate the importance of cache control mechanisms in rapidly drawing an entire field of view for rapid slide movements common in anatomic pathology. These findings are further explained in the discussion section.

## Discussion

Iris Core is designed to operate at the maximal performance available with modern hardware and was coded using an extremely granular graphical API to allow for optimizations specific to the unusual demands of real-time high data volume buffering and rendering imposed by WSV in a fully digital pathology department. A rendering engine must be able to efficiently buffer and render extremely large spatial data in real-time and with diagnostic image quality—requirements that are not present in other industries. This communication does not attempt to address the networking limitations or data storage limitations that hamper the adoption of digital pathology solutions; these topics are areas of future work and separate Iris modules. The performance metrics addressed in this communication were made on local WSI files to the Iris implementations to isolate and interrogate the rendering engine's performance.

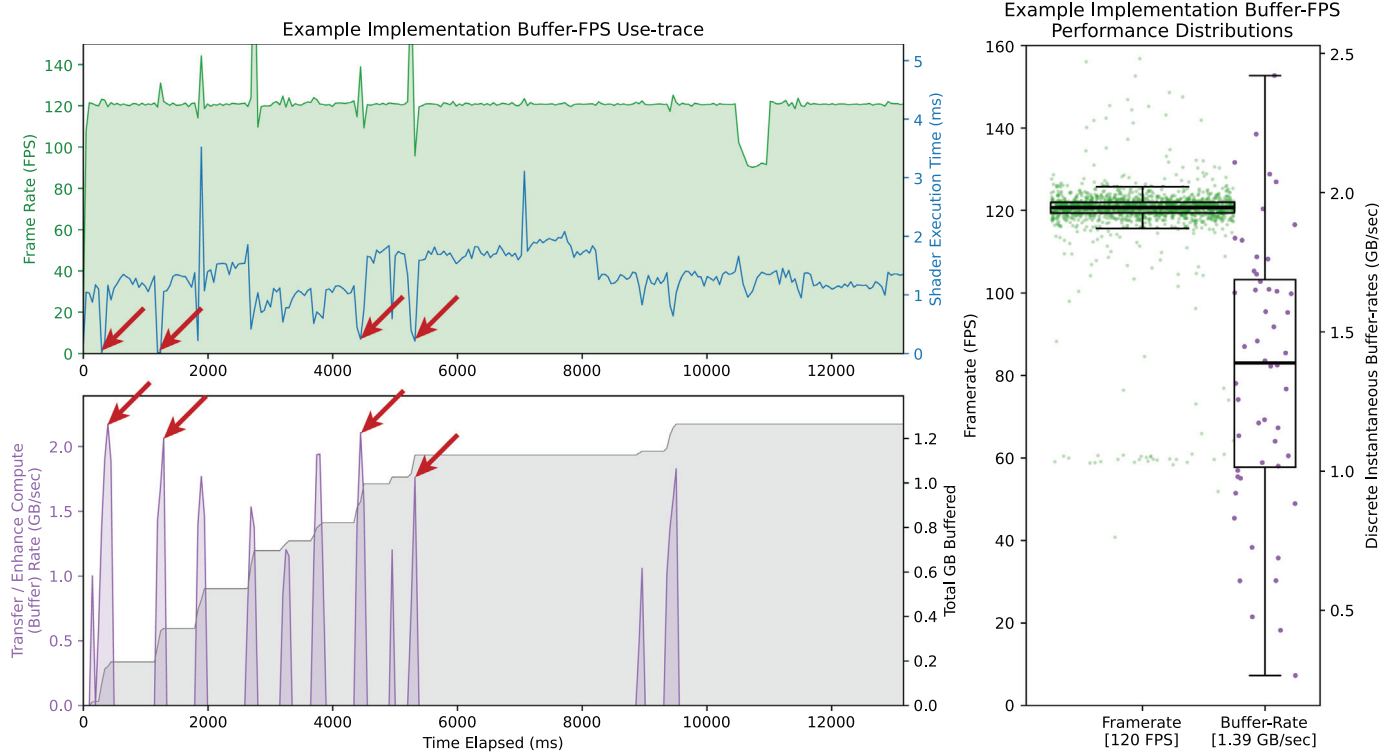

**Fig. 8.** Iris system performance trace for Example iPadOS implementation. Use-traces (left) over a 12 s period (approximately 7–9 ms smoothed over 48 ms) and corresponding performance distributions for the trace (right) showed sustained 120 FPS (119.4–121.9 FPS) rendering rate and ≤2 ms shader execution time with rapid slide navigation during buffering. Rapid tile buffer-rate (left-bottom) during the median 25 ms TeFOVs (Fig. 5) occurred at a median rate of 1.39 GB/s (1.01–1.69 GB/s). Shortened shader execution times were noted coincidentally with large data boluses (red arrows) and represent the high-speed access to the slide image data residing within the L1/L2 cache as a direct result of the preceding buffering transaction, without the overhead of pulling the image data from VRAM. This corresponds with increased FPS. Differences between the trace (left) and performance distributions (right) are due to averaged smoothing in the FPS/buffering trace. A drop in FPS was noted during Iris' interaction with the iOS pencil kit (external drawing calls) during an annotation event (short reduction to 80 FPS on right-side of trace at approximately 10.5 s). A video recording of use during the capture of this trace is provided as Video 1 in the online supplements.[19]

Iris Core is written based on the Vulkan API for several reasons. The most significant of these include the low-level/low-overhead nature of the API, which allows for the custom design of the engine to include optimizations specific to the rendering of WSIs, and for integration of compute shaders, which allow for the implementation of data analysis and transforms (such as AI) directly on the GPU-local image data. This includes operating on framebuffers even as they are rendered to the screen. Alternative engines and graphical API (such as OpenGL/WebGL) are less explicit and more dynamic; however, this dynamism comes with generic execution models that are not fine-tuned to the unique requirements of WSI rendering, described above. By writing Iris Core's engine with Vulkan, the graphical pipelines are more explicit and complex but the granular characteristics, including GPU cache synchronization, can be explicitly tailored to this unique use of graphical API.

Iris Core demonstrates extremely high performance in both evaluated criterion: rendering speed and data buffer-rate. The rendering speed shows sustained maximal frame-rate on all test platforms. Frame-rate variation in Iris corresponds with large buffering transactions. Shader execution time showed a transient decrease with large buffering transactions (Fig. 8, *red arrows*), which may seem unexpected. High GPU demand would be expected to consume a greater share of the GPU's capacity and thus prolong the rendering shader execution time; paradoxically, the execution time dramatically decreases. Reasons underlying this include Iris' GPU cache control mechanisms with separate queue families for buffering and rendering execution. The GPU cache transfer operations allow the rendering pipelines high-speed access to slide image data already residing within the L2 cache. These data are available as a direct result of the preceding buffering transaction on a separate queue. The subsequent draw command for the view no longer requires the overhead of pulling relevant image data into the GPU cache from VRAM; rather, it uses the already available data from the separate buffer queue transactions. The rendering shader can then simply render. When moving into an entirely new field of view, the rendering shader execution time approaches zero milliseconds and exemplifies the importance of the granular control Iris can achieve.

The data buffering rate metrics demonstrate the capacity to render an entirely new field of view in less time than is required for a standard monitor to refresh and fill in a HR high-quality view in only two to three frames. A standard monitor has a display refresh of 60 Hz, or an average frame time of 17 ms, whereas Iris achieves a median LR-TeFOV of 10 ms with an HR-TeFOV of 30 ms when using Iris Codec to decode the image files. This means that a new scope field is immediately available when shifting the entire field of view, achieving performance like that of a glass slide. This comparison with glass, and most importantly the types of saccadic slide movements made by anatomic pathologists, is among the reasons we are explicit in our data buffer-rate criterion—specifically that a FOV in the TeFOV metric must have no pixel overlap.

Despite being the most relevant metric to the practicing pathologist, the TeFOV can vary greatly with the current objective, zoom amount, and window resolution.[7,8,20] This makes the TeFOV an inconsistent choice when evaluating performance and should be used together with TPT as a standard performance pair. The TPT is normalized with respect to the number of tiles per FOV transaction. The reciprocal can be used to calculate bitrate as the product of the bytes per tile (in our above result, an 8-bit four-channel or 32-bit pixel with $256 \times 256$-pixel tile). Bitrate is even more generalized for comparisons across platforms. Iris can decode an image tile, transfer it to GPU-local VRAM, and compute a reduction-enhancement floating point kernel on multiple simultaneous downsampled levels all with a median duration of less than 89 μs per tile for HR layer. These metrics and capabilities far exceed what have previously been reported by up to an order of magnitude (10–25 ms/FOV) in TFOV relative to the most performance intensive contemporary works of DziTileSource (357 ms/FOV),[4] FlexTileSource(240 ms/FOV),[8] and further decoder optimizations done by FlexTileSource authors Schuffler *et al.* (164 ms/FOV).[3] These TFOV metrics are not as strict as TeFOV requirements and do not include reduction-enhancement by integrated compute pipelines. The more normalized TPT values were also over an order of magnitude faster (100–160 μs/tile) than the maximum literature-reported values using Aperio WebViewer ($\geq 2000$ μs/tile, reported as $\leq 500$ tiles/s).[7] The order of magnitude performance gains achieved by shifting to a new form of render engine rather than tile size optimizations of thin client Javascript based rendering solutions illustrates the important paradigm shift of adopting new render engine paradigms based upon modern graphical APIs to overcome the inherent and usually insurmountable limitations of prior rendering technologies.

For digital pathology to effectively transition from being merely a supplementary tool to a primary diagnostic modality, the following enabling conditions will be essential.

Adoption of modern frameworks: Transitioning from antiquated JavaScript-based viewers to frameworks optimized for GPU and CPU acceleration will be critical. This could involve new software entirely or significant enhancements to existing ones. Iris is one such example of a GPU-optimized framework.

Investment in high-performance computing infrastructure: Healthcare institutions must invest in modern computing infrastructure that includes high-end CPUs, GPUs, and networking capabilities to support real-time, large-scale image processing, and on GPU compression systems (future Iris modules).

Integration with AI and machine learning: Combining hardware acceleration with advanced machine learning models can leverage GPUs to process vast amounts of data quickly, aiding in identifying patterns and anomalies that may be time intensive or difficult for the human eye to detect. Iris can directly integrate GPU compute modules and blend rendering with real-time on-GPU analysis.

Integrating advanced CPU and GPU technologies into digital pathology holds immense promise for transforming primary diagnosis. By overcoming the limitations of current WSV software and embracing super-pipeline architectures, the field can achieve unprecedented performance, scalability, and accuracy. A critical recipe for the successful development of these extremely high-performance tools requires a unique recipe for software development. This includes creation or significant involvement by pathologists who have a profound understanding of the pragmatic implications introduced by digital sign-out. It also involves development by programmers who share a deep understanding of the hardware capabilities and thus who can solve the complex anomalies in rendering behaviors intrinsic to extreme levels of high-data throughput parallel programming. The development of Iris accomplishes both of these requirements. As the healthcare industry evolves, these technological advancements will allow digital pathology to flourish as a pivotal diagnostic tool, ultimately improving patient outcomes through enhanced diagnostic capabilities. We believe that Iris represents a fully mature exemplar of what is possible when CPU and GPU architectures are utilized to their full potential.

## Author contributions

RL designed and programmed the Iris platform, developed the performance testing, wrote elements of the manuscript, and created the manuscript figs. UB provided guidance, mentoring, and wrote elements of the manuscript. All authors read and approved the final article.

## Data availability statement

## Declaration of competing interest

The authors declare the following financial interests/personal relationships which may be considered as potential competing interests:

Ryan Landvater has patent #US20230334621A1 pending to REGENTS OF THE UNIVERSITY OF MICHIGAN. If there are other authors, they declare that they have no known competing financial interests or personal relationships that could have appeared to influence the work reported in this article.

## Acknowledgements

We have no additional funding sources to disclose. We would like to acknowledge Drs. Michael Olp, Corey Post, and Vincent Laufer for manuscript review. We would like to acknowledge the work of Drew Bennett, Director of Software, Content Licensing, and Research Partnerships at the University of Michigan. His assistance was instrumental in the development of Iris. We would like to acknowledge the authors and maintainers of the OpenSlide, vendor and system agnostic slide tile decoder, including Benjamin Gilbert, for their wide contribution to the field. Iris uses OpenSlide in the Iris Codec encoder and can use it during slide viewing for vendor files. We would like to acknowledge Vinnie Falco, author of many Boost C + + open-source libraries, including Boost Beast, a high-performance WebSocket library. Vinnie provided personal responses to aid in debugging networking performance issues and explanations of networking concepts critical to performance. We would like to acknowledge Kenton Varda, the author of Protobuffers and Cap'n Proto serialization libraries. His responses were valuable in the design of the Iris Codec file format. Further, methodology in Iris' serialization routines is loosely based upon Kenton's and Google's Flatbuffer serialization code base. We would like to acknowledge the work performed by the Kronos Group in the development of Vulkan and the team behind MoltenVK.

## Appendix A. Glossary of Terms

**atomic**: A computer operation which cannot be (or is not) interrupted by concurrent operations. This type of computer programming operation must complete before any other concurrent process can access or modify the underlying variable data.

**blocking**: Stopping the execution of the current process to wait until some condition (such as a needed variable becomes available or a signal is received) has been met.

**buffer-time**: Data transfer rate metric describing the duration the buffer thread spends transferring image tile data for a full field of view. This comprises either the high-resolution or low-resolution time for field of view (TFOV).

**buffer-rate**: Data transfer rate of the buffer thread transferring image tile data and downsampling that data for high-quality rendering in terms of raw bytes (gigabytes) per second.

**codec**: A hardware or software-based process that compresses and decompresses large amounts of data. In this case, a codec is the underlying software that compresses or decompresses slide image data.

**convolution**: A linear algebra operation used in visual processing and signal processing. With specific kernel functions, it is used for blurring, sharpening, edge detection, and an almost infinite array of mathematical transformations. Mathematically, the convolution operator involves the integral of products between the two functions $f$ and $g$ as follows:

$$\int_{-\infty}^{\infty} f(\tau) * g(\tau - t)d\tau \tag{A.1}$$

where $t$ is the start location in the dataset (the pixel offset of the tile) and $\tau$ is the size of the kernel, which changes based upon the degree of downsampling for the reduction-enhancement pipeline.

**Laplacian (of Gaussian) kernel:** Also known as the Ricker Wavelet or "the Mexican hat wavelet", due to taking the shape of a sombrero. It is the negative normalized second derivative of a Gaussian function. It can be used for sharpening transforms or edge detection based upon the normalization of the wavelet integral. The 2D expression in x and y dimensions can be represented as:

$$\psi(x, y) = \frac{1}{\pi \sigma^4}\left(1 - 0.5 * \left(\frac{x^2 + y^2}{\sigma^2}\right)\right)e^{-\frac{x^2+y^2}{2\sigma^2}} \tag{A.2}$$

where sigma represents the scale.

**mipmap:** Pre-calculated progressively smaller versions of an original image (downsampled by $2\times$, $4\times$, $8\times$, etc.) that are rendered to reduce visual artifacts such as blocking artifact or Moiré patterns.[15]

**pipeline:** Also known as a data pipeline, it is a set of computer processing elements (parts of a computer program) connected in series, where the output of one element is the input of the next one. The elements of a pipeline are often executed in parallel or in time-sliced fashion.[12]

**shader:** A small computer program written for execution on a GPU rather than a CPU. These programs are specially designed to be run in parallel to maximize the amount of data that can be computed via SIMD instructions (*see below*) all at once.

**SIMD** (Single Instruction Multiple Data): A computer instruction that evaluates a single mathematical expression on multiple independent data simultaneously. Example: Instead of performing $A + C = AC$ sequentially followed by $B + D = BD$, the computer can compute $[A, B] + [C, D] = [AC, BD]$ in a single use of the computer processor. This is the major way by which the GPU achieves massive parallel computation.

**thread:** An execution sequence of computer code. A thread is the smallest part of a computer process that includes local variables. A processor switches between threads in the execution of parallel code and multiple computer cores execute threads simultaneously.

**thread-safe:** An assurance by the developing computer programmer that data variables within the API will not cause programmatic bugs if two threads simultaneously attempt to modify the same data or perform the same instructions on shared resources.

**tile-transfer rate:** Data transfer rate of the buffer thread transferring image tile data and downsampling that data for high-quality rendering in terms of the number of $256 \times 256$ pixel tiles per second. This value is the reciprocal of the time per tile metric.

**wrapper:** A computer programming technique of embedding some underlying computer programming functionality within a computer programming system with additional functionality. An example would be to consider a computer programming object "horse" and a wrapper object "horse-with-saddle". The wrapper refers to the horse but added the

"with-saddle" feature. The underlying computer memory in Iris, a part of the operating system, has added atomic-signaling functionality wrapped around it to allow for synchronization.

Supplementary data to this article can be found online at https://doi.org/10.1016/j.jpi.2024.100414.